**The multiplicity of memory enhancement: Practical and ethical implications of the diverse neural substrates underlying human memory systems**

Kieran C. R. Fox[a]*, Nicholas S. Fitz[b], and Peter B. Reiner[b]

[a] Department of Psychology, University of British Columbia, 2136 West Mall, Vancouver, B.C., V6T 1Z4 Canada

[b] National Core for Neuroethics, Department of Psychiatry, University of British Columbia, 2211 Wesbrook Mall, Vancouver, B.C., V6J 2B5 Canada

* To whom correspondence may be addressed (at address [a] above). Telephone: 1-778-968-3334; Fax: 1-604-822-6923; E-mail: kfox@psych.ubc.ca

# The multiplicity of memory enhancement

**Abstract**

The neural basis of human memory is incredibly complex. We argue that the diversity of neural systems underlying various forms of memory suggests that any discussion of enhancing 'memory' per se is too broad, thus obfuscating the biopolitical debate about human enhancement. Memory can be differentiated into at least four major (and several minor) systems with largely dissociable (i.e., non-overlapping) neural substrates. We outline each system, and discuss both the practical and the ethical implications of these diverse neural substrates. In practice, distinct neural bases imply the possibility, and likely the necessity, of specific approaches for the safe and effective enhancement of various memory systems. In the debate over the ethical and social implications of enhancement technologies, this fine-grained perspective clarifies – and may partially mitigate – certain common concerns in enhancement debates, including issues related to safety, fairness, coercion, and authenticity. While many researchers certainly appreciate the neurobiological complexity of memory, the political debate tends to revolve around a monolithic one-size-fits-all conception. The overall project – exploring how human enhancement technologies affect society – stands to benefit from a deeper appreciation of memory's neurobiological diversity.

## 1. The imperfections of memory

"To err is human," wrote Alexander Pope (1711). Yet we are confronted so directly with our own forgetfulness (Ponds, Commissaris, & Jolles, 1997), with faulty memory leading to distorted eyewitness testimonies (Loftus, 2013), and increasingly, with the tragic decline of memory in neurodegenerative conditions such as Alzheimer's disease (Reitz & Mayeux, 2014), that the suboptimal functioning of human memory systems practically cries out for correction. Whereas in evolutionary terms the limitations of our memories might represent a subtle and intricate balancing act where apparent drawbacks are ultimately beneficial or compensated for in other ways (Schacter, 2002), people appear nonetheless to be increasingly unsatisfied with this rich natural endowment. Even the most gifted among us can only keep so much in mind at a given time; can only remember so far back into the past with clarity; can only master so many skills; can only recall so many facts at the right time – and both individuals and private companies are moving rapidly to amend these perceived deficits by artificial means. Moreover, even as individuals want and expect to live longer, healthier lives with their memories and skills intact, modern societies and educational systems are creating new demands on our memory that our natural neurobiology is (arguably) unequipped to handle.

For the purposes of individuals living in modern technological societies, then, memory is so imperfect, and its proper functioning so crucial to nearly every aspect of human behavior, that it is no surprise that it plays a leading role in the burgeoning debate on cognitive enhancement (Bostrom & Sandberg, 2009; Cabrera, 2011; Farah et al., 2004; Greely et al., 2008; Hildt & Franke, 2013; Hyman, 2011; Liao & Sandberg, 2008). While some authors have drawn attention to different memory systems in their discussions (Cabrera, 2011; Liao & Sandberg, 2008), to our knowledge there has yet to be an in-depth exploration of how the existence of various memory systems has direct implications for both practical and ethical concerns in memory enhancement. The central aim of this paper is to explore these issues. We believe such an exploration is important, because the lack of consideration typically given to the subtleties and distinct systems of memory in humans tends to result in both unrealistic enthusiasms and objections in the debate on enhancing or diminishing memory with emerging technologies. For instance, Sandberg (2011) casually claims that researchers have discovered "genes in humans whose variations account for up to 5% of memory performance" (p. 76), without discussing which form (or forms) of memory the researchers investigated. In fact, the original study investigated both episodic and working memory and found that the genetic variations identified had no relationship to working memory ability (de Quervain & Papassotiropoulos, 2006), illustrating our point that different forms of memory will require understanding multiple biological substrates. Even major, high-profile reviews of cognitive enhancement have tended to discuss 'memory' in a relatively undifferentiated way, focusing on episodic memory without calling it such, and discussing working memory instead in a section on executive function (Farah et al., 2004).

Beyond this general lack of specificity in discussing 'memory' enhancement or modification, some of the most compelling concerns, e.g. with respect to fairness or authenticity, can be relatively easily seen to be specific to certain memory systems much more than others. For instance, an important and oft-echoed concern is that if memory enhancement is costly and therefore restricted to the upper classes of society, it might lead to unfairness and greater societal inequality (Fukuyama, 2003; Savulescu, 2006). This concern has obvious importance for forms of memory that can be used as competitive skills, such as semantic memory and procedural (skill) memory – but it is much more difficult to see how enhancement of individual episodic memories could lead to any competitive edge.

The multiplicity of memory enhancement

The purpose of this paper is therefore twofold: first, to highlight the conceptual and neurobiological differences between memory systems and insist that talking about 'memory' in general is too imprecise to be helpful in such a discussion; and second, to point out how common practical and ethical concerns in the debates surrounding memory enhancement specifically, and cognitive enhancement more broadly, often apply only (or at least largely) to certain types of memory. Philosophers and others have argued for some time that the idea of a monolithic 'memory' as a natural kind is suspect for many reasons (Cheng & Werning, 2016; Michaelian, 2011, 2015; Rupert, 2013); we argue that the dissociable neural bases of various memory systems corroborate these criticisms, and portend the possibility of, and perhaps need for, multiple biotechnological pathways to mnemonic enhancement. Moreover, these dissociable neural substrates benefit discussion of the ethical and social implications of enhancement, which can be made more concrete by reference to particular types of memory and their neurobiological corollaries. Indeed, the dissociable nature of memory systems bears on a range of concerns, including worries about safety, fairness and distributive justice, authenticity, and social pressure to enhance once a heightened level of ability becomes the 'new normal' (Cakic, 2009; Chatterjee, 2004; Farah et al., 2004; Fitz, Nadler, Manogaran, Chong, & Reiner, 2014; Greely et al., 2008).

The aim of the paper is not therefore to raise *new* ethical or practical issues, but rather to highlight how a fine-grained neuroanatomical/neurobiological perspective can help address these common themes in the enhancement debate, and perhaps allay some common concerns as well. First, we provide an accessible introduction to the neurobiological dissociability of memory systems for neuroethicists, philosophers, and others interested in this debate. Second, we discuss the practical implications of these differences in neural substrate throughout the brain, commenting on the likely requirement of diverse biotechnological techniques for enhancement or diminishment of distinct memory systems (we focus mostly on enhancement here, as selective memory diminishment/erasure remains highly speculative, whereas many technologies and pharmaceuticals are already aiming at, or even claiming success with, enhancement). Third, we relate this discussion to the major social and ethical issues surrounding enhancement (e.g., safety, distributive justice, peer-pressure, identity, and authenticity). Finally, we close with a note on some acknowledged similarities between memory systems and some challenges to the very notion of multiple memory systems, so as not to overemphasize their differences, and we consider future directions in the memory enhancement debate.

## 2. The multiplicity of memory

Although memory systems appear to interact seamlessly in healthy people, research on patients with 'neuropsychological' brain lesions, experimental animal models, and noninvasive functional neuroimaging of humans reveals at least four major systems of memory (Eichenbaum & Cohen, 2001; McDonald & White, 1993; Poldrack & Packard, 2003; Squire, 2004; Squire, Knowlton, & Musen, 1993; Tulving, 1985) (see Fig. 1). These include (i) short-term *working* memory; (ii) *procedural* or skill memory; (iii) *episodic* or autobiographical memory of specific events and experiences; and (iv) general *semantic* memory for facts and concepts. There are also other subtypes that, for the sake of brevity, we do not discuss in detail, including priming memory (Tulving & Schacter, 1990); perceptual representation systems (Schacter, 1990; Tulving & Schacter, 1990); classical conditioning (Clark & Squire, 1998); habituation (Rankin et al., 2009); and sensitization (Ji, Kohno, Moore, & Woolf, 2003) (see Fig. 1). Here, we provide overviews of the four major memory types noted above, along with important (albeit necessarily simplified) indications of the neural substrates specific to each type.

*Working* memory is often likened to the mind's 'blackboard,' an incredibly flexible but short-term workspace that can accommodate information from other forms of memory, and all kinds of sensory content. Working memory involves the manipulation and evaluation of other memories and incoming sensory information, allowing for complex tasks such as language comprehension and reasoning (Baddeley, 1992, 2012). Working memory recruits a wide network of brain regions depending on the content being manipulated, but depends most critically on several regions of the prefrontal cortex (Baddeley, 2012); Fig. 2).

*Figure 1.*   The varieties of human memory and correspondingly diverse neural substrates.

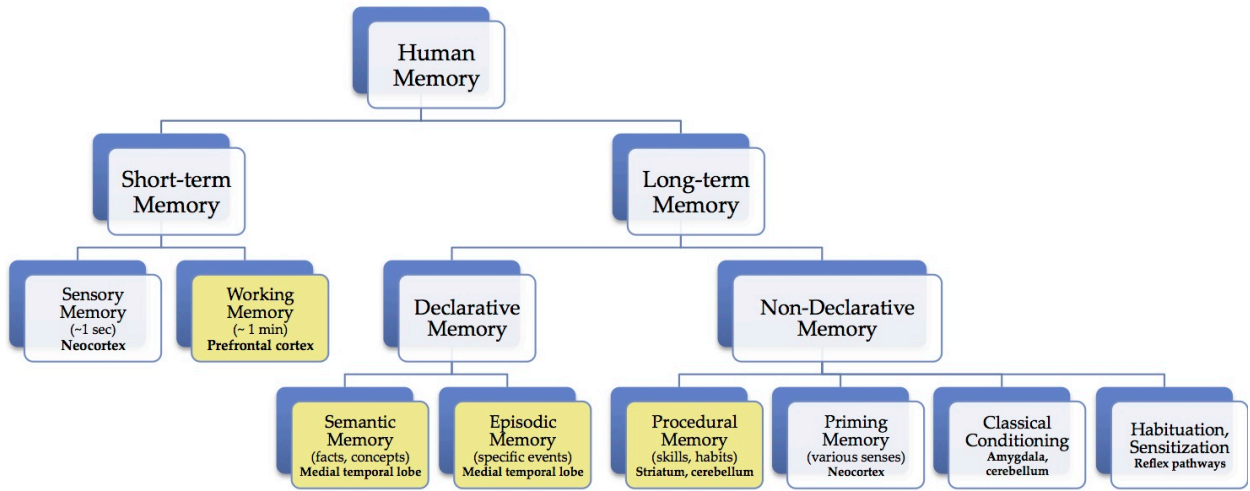

A simplified model of human memory shows several major systems that are distinguishable not only by their behavioral expression, but also by the brain structures critically involved (indicated in bold font for each major memory type). The four major systems highlighted in the present work are indicated in yellow. Following the model of Squire (2004).

*Procedural* memory broadly groups together many forms of 'implicit' skill learning – implicit in that generally we cannot consciously declare or explain the knowledge we have stored, despite being proficient at the skill itself (although some researchers have recently questioned this classic definition; cf. Stanley & Krakauer, 2013). Classic examples are learning how to play a musical instrument, learning to drive a car, or learning how to dance (Eichenbaum & Cohen, 2001; Squire, 2004). Such forms of memory appear to depend crucially on the striatum (a group of subcortical grey matter nuclei, also known as the *basal ganglia*) and the cerebellum (the 'little brain' tucked under the cerebral hemispheres and behind the brainstem; see Fig. 2).

*Semantic* memory is often that which we most take for granted: the enormous stockpile of facts, figures, and general information that we have stored up over a lifetime of experience. A classic example is the knowledge that Paris is the capital of France – in the absence of any memory or knowledge of when or where this fact was acquired. The hippocampus and adjacent medial temporal lobe structures (Fig. 2) are critical for the formation of new semantic memories (Eichenbaum & Cohen, 2001; Squire, 2004), but after a long enough time (most likely several years), semantic memories become independent of the medial temporal lobe and appear to become diffused throughout the neocortex (McGaugh, 2000; Moscovitch et al., 2005; Winocur & Moscovitch, 2011). Even neurological patients with complete bilateral loss of the hippocampus can recall semantic knowledge about themselves and the world, so long as this knowledge was acquired sufficiently long ago (several years, depending on different models and patients (Moscovitch et al., 2005; Squire, Stark, & Clark, 2004)).

*Episodic* or *autobiographical* memory comprises what is traditionally thought of as 'memory' by most people: the episodic recall of a specific event in a particular place and time (Conway & Pleydell-Pearce, 2000; Rubin, 2006; Thompson, Skowronski, Larsen, & Betz, 2013). Like semantic memory, the formation (i.e., encoding) of new episodic memories is critically dependent on the hippocampus and other medial temporal lobe structures (Fig. 2), but over time such recollections likewise become independent of the medial temporal lobe and appear to become dispersed throughout other neocortical brain areas (Moscovitch et al., 2005; Squire et al., 2004; Winocur & Moscovitch, 2011).

*Figure 2.*   Largely distinctive neural bases for diverse systems of human memory.

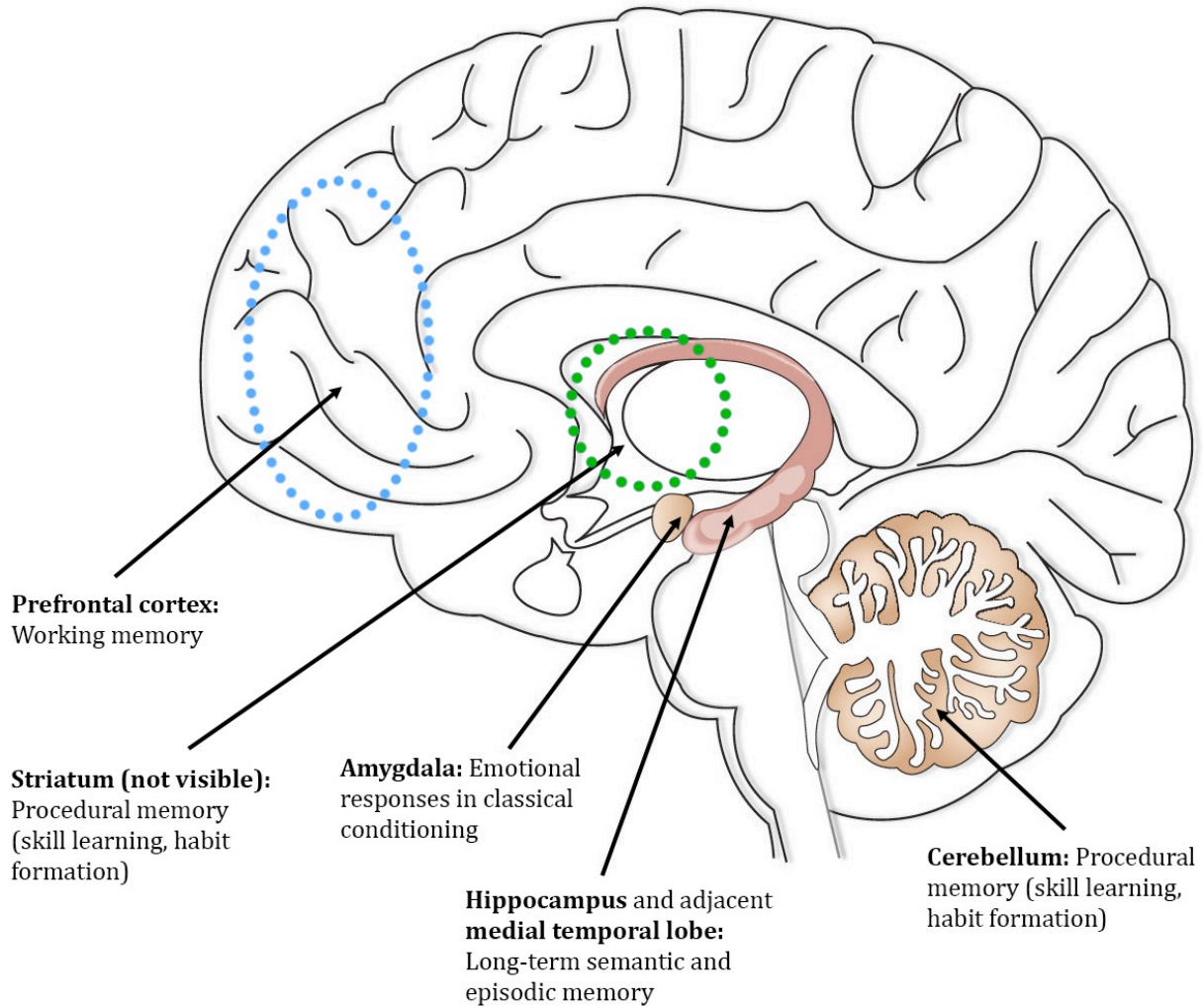

A highly simplified model of the diverse (and largely dissociable) neural substrates for different systems of human memory. In most cases, the structures indicated are crucial not only for the initial learning (encoding) of material, but are necessary for its later recall or expression. The exception appears to be the hippocampus and adjacent medial temporal lobe structures: although critical for the initial formation and shorter-term storage of episodic and semantic memory, such memories are somehow consolidated to other neocortical brain areas over time and eventually become independent of these structures. The consolidation process is inferred from patients with medial temporal lobe brain lesions, but its neurobiological mechanisms remain poorly understood. See the text for further details.

A highly simplified model of the various memory systems and their putative neural substrates is illustrated in Figure 2. Most striking is that both semantic and episodic memory appear largely dissociable from all other memory systems. In the clinic, patients with bilateral medial temporal lobe lesions can still be primed and conditioned, retain intact working memory, and can learn complex, novel motor skills such as 'mirror tracing' (Milner, Corkin, & Teuber, 1968; Squire, 2009). Yet these patients have severe 'anterograde' amnesia: that is, they are virtually unable to acquire *new* semantic or episodic memories. The literature from animal models further supports these dissociations (e.g., (Kesner, Bolland, & Dakis, 1993; Packard & McGaugh, 1992; Squire, 2004)), and subsequent research in humans with brain lesions has also shown striking double-dissociations. In one study, for example, medial temporal lobe patients with amnesia were directly compared to Parkinson's disease patients with severe damage to the dopaminergic neurons of the substantia nigra, which provides one of the major inputs to the striatum/basal ganglia (Knowlton, Mangels, & Squire, 1996). Patients with medial temporal lobe damage were (as expected) unable to recall their experience in a training session to learn a probabilistic classification task that required the formation of new habits, yet their learning of the task was normal. In contrast, the Parkinson's patients with severely compromised striatum/basal ganglia function recalled the training session without trouble, but failed to learn the relatively straightforward classification task (Knowlton et al., 1996).

Importantly, there is now extensive evidence that most of these memory types can be grouped into one of two umbrella categories: (i) *declarative* memory, which is consciously accessible and can be described and expressed voluntarily (in words, for instance); this category includes both semantic and episodic memory (Fig. 1); and (ii) *non-declarative* memory, which is not readily (or, at all) accessible to conscious awareness; rather, it constitutes implicit learning of skills, or tendencies of perception; it is "expressed through performance rather than recollection" (Squire, 2004). This category includes procedural memory for motor skills like learning an instrument, priming, habituation, and sensitization (Fig. 1). The support for this distinction, and the relevant point for the discussion here, is that these two forms of memory can be nearly completely dissociated at the neuroanatomical level. Indeed, neurological patients with circumscribed brain lesions in the medial temporal lobe (whose declarative memory is obliterated) do retain intact non-declarative forms of memory; extensive evidence from animal models further supports such a division (Milner et al., 1968; Schacter & Tulving, 1994; Scoville & Milner, 1957; Squire, 2004, 2009).

## 3. Practical considerations in the modification of dissociable memory systems

The naturally concerted action of the various systems described above should not blind us to the fact of their generally high level of dissociability at the neurobiological level. Although strict one-to-one correspondences between structures and functions should be neither sought nor expected in any object as interdependent as a nervous system (Cacioppo & Tassinary, 1990), nonetheless enormous progress has been made in the last few decades in delineating the necessary (if not necessarily sufficient) neural substrates for several different types of memory (expertly and thoroughly reviewed by (Eichenbaum & Cohen, 2001; Squire, 2004, 2009)). Here, we focus on a few of the practical implications of such neurobiological dissociations.

*Differences in physical location*

The simple fact that the *physical location* differs among these various brain structures has clear implications. Several putative enhancement techniques, including brain stimulation technologies – e.g., transcranial electrical stimulation (TES) and transcranial magnetic stimulation (TMS) – intervene above (and through) specific locations of the brain's cortical surface. Because these techniques require careful placement of electrodes at particular locations on the scalp, they may provide an opportunity to target specific memory systems. These facts have been exploited in studies aiming, for instance, to improve the acquisition and retention of motor skills by targeting the cerebellum (Ferrucci & Priori, 2014), to ameliorate the semantic and episodic memory deficits characteristic of Alzheimer's disease (Boggio et al., 2009; Ferrucci et al., 2008), and to enhance the reconsolidation of long-term semantic memory (Javadi & Cheng, 2013).

The same argument applies to invasive brain stimulation technologies such as deep brain stimulation (DBS), which involves semi-permanently implanting electrodes directly into deep brain tissue (Perlmutter & Mink, 2006). Although DBS has mostly been used as an experimental procedure to treat, or at least mitigate the symptoms of, severe psychiatric and neurological disorders such as depression and Parkinson's disease (Perlmutter & Mink, 2006), there have been efforts to use the technique to enhance memory (Hamani et al., 2008). Importantly, the DBS employed by Hamani and colleagues (2008) primarily drove increased activity in medial temporal lobe structures (see Fig. 2), and resulted in the selective enhancement of certain forms of memory, but not others. More recent studies have followed up on this notion of selective memory enhancement via DBS (Suthana et al., 2012), including for patients with diseases that severely affect memory, such as Alzheimer's (Laxton et al., 2010). Clearly, the physical location at which technologies such as brain stimulation are administered will be a critical factor in the success or failure of future interventions (Wang et al., 2014).

*Variations in neurochemistry*

Another practical issue to consider is that the brain is *neurochemically heterogeneous*, an observation first documented in the pioneering studies of Dahlström, Fuxe, and colleagues (Andén et al., 1966; Dahlström & Fuxe, 1964). The most salient illustration of this is that the type and frequency of neurotransmitters vary across different brain regions. Furthermore, these neurotransmitters interact with a range of receptor subtypes which are themselves heterogeneously distributed throughout the brain (Zilles & Amunts, 2009). For example, glutamate is the primary excitatory neurotransmitter in the central nervous system, yet there are multiple glutamate receptor subtypes that are differentially concentrated throughout, and even within, various brain regions. Far more marked differences between brain areas are evident when one begins to consider the profiles for multiple neurotransmitters (such as GABA and acetylcholine) simultaneously (Zilles & Amunts, 2009; Zilles et al., 2002). Such neurotransmitter receptor fingerprints are particularly relevant to any attempt to enhance memory using pharmacological means (Lynch, 2002; Lynch, Palmer, & Gall, 2011): they invite the development and refinement of receptor-subtype specific pharmaceuticals with preferential effects on brain regions with specific receptor fingerprints, and thereby, effects specific to certain forms of memory.

Certain interventions have already built upon this knowledge. For instance, the 'cholinergic hypothesis' posits that the severe declines in semantic and episodic memory characteristic of Alzheimer's disease are related to disruption of cholinergic neurotransmission, particularly in the hippocampus and surrounding medial temporal lobe tissue (Bartus, Dean, Beer, & Lippa, 1982; Francis, Palmer, Snape, & Wilcock, 1999). This has led to the clinical development of a series of drugs that inhibit acetylcholinesterase (the enzyme that breaks down acetylcholine at the synapse) in an effort to heighten the neurotransmitter's signaling by increasing its availability at the synapse (Corey-Bloom, 2003). Such agents are among the most commonly prescribed drugs for the treatment of Alzheimer's disease and other dementias (Hansen et al., 2008; Lanctôt et al., 2003; Tariot et al., 2000; Trinh, Hoblyn, Mohanty, & Yaffe, 2003), although their effectiveness for enhancement of individuals with normal cognitive function remains controversial (Repantis, Laisney, & Heuser, 2010; Rokem & Silver, 2010).

*Do enhancements also entail trade-offs?*

A third practical consideration focuses upon trade-offs between different forms of memory. The brain is an interdependent system (Bassett & Bullmore, 2006) with its own homeostatic regulatory features (Cacioppo & Berntson, 2011; Weiss, Miller, Cazaubon, & Couraud, 2009), and it is quite plausible that improving performance in one domain might adversely affect another (Hills & Hertwig, 2011; Maslen, Faulmüller, & Savulescu, 2014) – sometimes referred to as the *net zero-sum* problem (Brem, Fried, Horvath, Robertson, & Pascual-Leone, 2014). This could result, for instance, from finite resources (e.g., glucose, oxygen, or precursor molecules for the synthesis of neurotransmitters) being diverted from certain areas to others.

The few studies that have investigated this issue seem to indicate that sometimes trade-offs exist and other times they do not. One study tested the idea that Adderall (amphetamine), which tends to increase focused, sustained attention, might have detrimental effects on creative forms of thinking, which tend to require a broader, more dispersed form of attention for successful execution (Farah, Haimm, Sankoorikal, & Chatterjee, 2009). The authors found no evidence of such an antagonistic effect for this particular drug across these particular domains. Other research , however, *has* found evidence for trade-offs in the use of TES (Iuculano & Kadosh, 2013; Sarkar, Dowker, & Kadosh, 2014). Iuculano and Kadosh (2013), for instance, found during a mathematical learning task that stimulation of posterior cingulate cortex enhanced numerical learning, but simultaneously impaired automaticity for the newly learned material. Conversely, stimulation of dorsolateral prefrontal cortex had the opposite pattern of effects, resulting in impaired learning but enhanced automaticity. The question of whether such trade-offs are inevitable, occasional, or perhaps actually spurious, can only be answered by further research investigating these questions in detail for various memory systems (Kadosh, 2015; Maslen et al., 2014).

**4. Ethical and social implications of distinctive memory systems**

The discussion around the ethical and social implications of enhancing or diminishing memory raises at least four cardinal concerns (Chatterjee, 2004; Fitz et al., 2014; Greely et al., 2008; Hildt & Franke, 2013). Most prominently, these issues include:

- Safety, including side-effects and the balance between risks and benefits
- Social pressure and coercion, explicitly from peers and implicitly from society
- Distributive justice
- Authenticity, cheating and identity

The observation that there are largely dissociable neural bases of the various memory systems raises an important question: do these ethical issues affect all memory systems equally? We suggest that in some instances the worries are identical across memory systems while in others, there are clear differences. In the discussion that follows, we highlight examples of neuroethical concerns that differ and are similar across systems. Our treatment of the issue is not intended to be exhaustive but rather illustrative, and we invite our colleagues to consider not only other concerns that might differ by memory system, but also the full range of cognitive, emotional, and moral capacities under discussion in the field (Cabrera, Fitz, & Reiner, 2015).

*Safety*

In certain ways, it seems odd that concerns over safety would be a worry for neuroethicists. For safety is really a matter for the medical establishment, ensuring that treatments are safe for humans to use. But what is really meant by the safety concern is not whether a given treatment has traditional side-effects but rather that we need to weigh the risk/benefit ratio of a given treatment. This becomes a rather more challenging exercise when the benefit is viewed as an enhancement (in 'healthy populations) than when it is viewed as a therapeutic (in 'unhealthy' populations). In other words, the fundamental safety issue can be framed as asking how meaningful are the benefits, given the known (and potentially unknown) risks of a given enhancement technology?

Indeed, considering the risks of cognitive enhancement is relatively straightforward and is often described in terms of the *mode* by which enhancement is achieved. For example, so-called 'non-invasive' brain stimulation (Davis & van Koningsbruggen, 2013) appears to involve relatively little in the way of risk, pharmacological agents a bit more, and deep brain stimulation even more risk. This spectrum of risk provides an opportunity for a simple calculus: for a given benefit, the lowest risk method is preferred.

But focusing upon risks alone does not provide perspective on the relative benefits of enhancing (or diminishing) different forms of memory. If we consider the four memory systems reviewed above (*working*, *procedural*, *episodic* and *semantic* memory), we can imagine that a small increase (for the purpose of this discussion, let's arbitrarily set it to 5%) in the function of each might provide quite different benefits. It seems likely that the most dramatic benefit would accrue from a small increase in working memory, as this is so fundamental to our ability to juggle concepts in our brains – allowing for the sort of synthetic analyses at which our brains excel. We might also imagine that enhancing episodic memory – enhancing, for instance, the vividness of a particular time in one's life – might be useful, as it relates to our historical conception of who we are as humans. But more exotic enhancements of episodic memory have already been envisioned: dream memory is most akin to episodic memory, and dream enthusiasts have begun using a variety of over-the-counter pharmaceuticals to enhance access to autobiographical memory within the dream, as well as subsequent memory of the dream experiences (upon awakening). These do-it-yourself experiments include the off-label use of acetylcholinesterase inhibitors designed specifically to combat declines in declarative memory in Alzheimer's disease and dementia (Yuschak, 2006, 2007). So, whereas enhancing episodic memory might be of comparatively little importance to some users, the example of dream enthusiasts highlights how small subcultures might value particular forms of memory enhancement very highly. Along these lines, improving procedural memory might be of particular utility and interest to professional athletes and musicians, but perhaps less desirable to many others. Finally, in the modern world with devices near at hand that readily provide encyclopedic knowledge at the touch of a button, the benefit of enhancing semantic memory seems less profound. We would be well served to supplement these speculative thoughts on the relative benefits of enhancing differing modalities of memory by empirical studies that compared the relative benefit that the public perceives improving each subtype of memory might afford.

A different way of thinking about the ethical implications of benefit versus risk arises when one considers therapeutic forgetting - the attempt to erase traumatic autobiographical/ episodic memories (Kolber, 2006). The therapeutic objective is certainly worthwhile: post-traumatic stress disorder is devastating to those who suffer from its effects and there is little controversy over whether one should provide relief to such individuals. More contentious is the notion of proactive treatment of individuals with agents that induce therapeutic forgetting – i.e., immediately treating victims of traumatic or severely aversive experiences with memory-blocking or affect-blunting drugs, even before there is any clear indication whether the experience would engender lasting negative effects or not. For traumatic experiences by no means *necessarily* lead to post-traumatic stress disorder, and coping with (and overcoming) a severely negative or traumatic experience can lead to personal growth and change (Adler & Hershfield, 2012; Bonanno, 2004; Evers, 2007). Indeed, viewed in this light, not treating an individual may be viewed as a benefit.

Different drugs will also have varying addictive potential. Caffeine, for instance – probably the world's most-used 'enhancement' drug today – is clearly addictive (Griffiths & Woodson, 1988; Juliano & Griffiths, 2004), and widely-used amphetamines such as Adderall also show considerable potential for both psychological and physical dependence (Nutt, King, Saulsbury, & Blakemore, 2007). Other enhancement drugs will also likely have addictive potential, and this potential may vary in relation to the memory system in question: drugs affecting the striatum/basal ganglia motor skill learning systems (Fig. 1), for instance, may be particularly risky in this regard. The basal ganglia are closely intertwined with a general dopaminergic 'reward' system (Schultz, Tremblay, & Hollerman, 2000) that is implicated in a host of addictions, including not just abuse of stimulants like cocaine and amphetamines (Grace, 1995), but also behaviors such as gambling and problematic/addictive internet use (M. Potenza, 2015; M. N. Potenza et al., 2003; Tamminga & Nestler, 2006). Drugs targeting basal ganglia structures or dopaminergic neurotransmission in general clearly have a high potential for abuse. Conversely, enhancement drugs targeting other brain areas or neurotransmitter systems, such as modafinil (Minzenberg & Carter, 2008; Repantis, Schlattmann, Laisney, & Heuser, 2010) may be much less liable to such problems.

*Peer-pressure and coercion*

The possibility that widely available enhancements might lead to peer- or employer-pressure (or even outright coercion) to enhance oneself is a very real concern (Vincent, 2013). Indeed, it seems undeniable that this kind of peer-pressure is already present in domains outside the enhancement of memory, such as in performance-enhancing drug-use by professional and amateur athletes (Savulescu & Foddy, 2011), or in the widespread use of amphetamines and other putative 'enhancements' in the military (Giordano, 2014; Greely, 2011; Lin, Mehlman, Abney, & Galliott, 2014; Moreno, 2003, 2006).

It seems likely that the advent of effective memory enhancement techniques will lead to similar problems for various forms of memory. But it is *unlikely* these problems will vex all memory systems equally. For instance, there are obvious competitive advantages to enhancing semantic, working, and procedural memory, in academic, medical, athletic, military, and various other settings (Appel, 2008; Beddington et al., 2008; de Sio, Faulmüller, & Vincent, 2014; Goold & Maslen, 2014). On the other hand, it is difficult to imagine a situation in which an employer (for instance) might coerce an employee into enhancing autobiographical memories of purely personal events. Similarly, peer-pressure is unlikely to come to bear on autobiographical memory for the same reasons: enhancement of personal memories is unlikely to be an important factor in keeping up with peers in competitive academic or athletic settings, for example.

*Distributive justice*

A main source of the concern over 'fairness' relates to the anticipated high costs of enhancement technologies (Fukuyama, 2003; Savulescu, 2006). If enhancements are expensive and confer significant competitive or personal advantages, there is a strong chance that they will be available only to the wealthy – and this lack of equal access to enhancements could reinforce current inequality, further broadening the gap between the haves and the have-nots (Fukuyama, 2003; Savulescu, 2006). Of course, our society is already full of such inequities (Norton & Ariely, 2011), and few would argue for restricting advances in healthcare or quality of life because of the potential for inequitable distribution. Unequal access is generally not grounds for prohibiting neurocognitive enhancement, any more than it is grounds for prohibiting other types of enhancement such as private tutoring or cosmetic surgery that are enjoyed mainly by the wealthy. Indeed, it might be the case that neuroenhancement, if financially affordable and available to all, could potentially help equalize opportunities in our society (Savulescu, 2009).

The diverse neural substrates of memory systems bear on these concerns. Although its efficacy remains unproven as a bona fide enhancer (Tremblay et al., 2014), there is considerable interest in the use of tDCS to enhance cognitive function, especially as the overall cost is within the realm of other consumer products. Thus, enhancing working memory, which is primarily reliant on superficial cortical areas on the brain's surface (see Fig. 2), may be possible with this more cost-effective form of enhancement, while there is less likelihood that tDCS would be effective in modulating deeper memory structures (Jie, Tiecheng, Yan, Duc, & Xiaoping, 2013) such as the striatum and medial temporal lobe, affecting procedural, semantic or episodic memories (see Fig. 2). In contrast, enhancement of these forms of memory might require more expensive interventions, such as pharmaceutical agents and DBS – therefore making these forms of memory potentially less likely to be fairly distributed to all.

Indeed, pharmaceuticals such as the acetylcholinesterase inhibitor galantamine have already been developed in an attempt to combat declines in episodic and semantic of memory (e.g., in patients with dementia or Alzheimer's disease; (Hansen et al., 2008; Tariot et al., 2000). They remain prohibitively expensive, however, and require continuous use in order to be even minimally effective (Loveman et al., 2006). Similarly, as discussed above, deep brain stimulation (DBS) has been used to attempt to enhance these declarative forms of memory via stimulation of medial temporal lobe regions (Suthana et al., 2012), but remains extremely costly – anywhere between $35,000 to $100,000 for a typical depth electrode montage (Fraix et al., 2006; Green, Joint, Sethi, Bain, & Aziz, 2004). These more expensive forms of (putative) cognitive enhancement are examples of the validity of the distributive justice argument.

*Identity, authenticity and related concerns*

Another central concern is that memory enhancement might threaten one's sense of identity (Cabrera, 2011; Erler, 2011). Prima facie, it seems that different forms of memory would be viewed differently in this kind of debate. It is easy enough to see how enhancing semantic memory, for instance, could be seen as *unfair* in competitive contexts, but it is much more difficult to see how a faster or greater accumulation of facts could fundamentally affect one's sense of *identity*. Likewise for enhancements of procedural memory (motor skills) – learning to play the piano or juggle more quickly and easily than you might have otherwise seems much less likely to entail any deep existential anxieties than, for instance, the selective erasure or dampening of autobiographical memories (Earp et al., 2014).

In contrast, it is hard to see any *competitive* advantage to enhancing autobiographical memory, which relates only to one's personal past. Beyond individual enjoyment or enrichment, there do not appear to be any personal or societal advantages to the enhancement of self-specific episodic memories. But enhancements or diminishments of autobiographical memory could profoundly affect one's sense of identity. What one chooses to remember (or forget) about oneself, what events are made more vivid, or conversely, dampened – such issues raise serious philosophical and ethical concerns (assuming, of course, that technologies allowing this kind of precise control eventually become available).

In fact, examples of enhancement of autobiographical memory have often served as the central objections to memory enhancement writ large (e.g., (Donovan, 2010; Earp et al., 2014; Henry, Fishman, & Youngner, 2007)). For instance, two concerns are that either (i) enhancement of 'memory' in general may lead to accidental 'spillover' enhancement of traumatic experiences such as rape or wartime trauma (Earp et al., 2014); or (ii) that the understandable desire to dampen such traumatic memories with pharmacological agents like propranolol will lead to 'over-medicalization' (Conrad, 2008; Elliott, 2004; Frances, 2013) of unpleasant memory, and an abuse of memory diminishment drugs by normal people without trauma or memories 'severe enough' to warrant such pharmacological interventions (Donovan, 2010; Henry et al., 2007). Unfortunately, semantic memory appears very closely tied to autobiographical memory at the neurobiological level ((Moscovitch et al., 2005; Squire, 2004; Squire et al., 2004; Winocur & Moscovitch, 2011); see also Fig. 2). The selective enhancement of semantic knowledge of facts and concepts, without a corresponding enhancement of autobiographical memory, therefore remains a difficult (perhaps even intractable) challenge for future research.

Similar to the 'fairness' critique that an artificially enhanced achievement is not entirely one's own (and is akin to 'cheating'), an artificially enhanced memory might suggest that the person carrying these enhanced memories is somehow less 'authentic' (Elliott, 1999). In contrast, others have argued that the 'original,' more-flawed self could just as easily be seen as the inauthentic one, and that therefore there is no need to view an enhanced self as lacking authenticity (Levy, 2007).

**5. Counterpoint: Similarities across memory systems and challenges to the systems view**

Although our central argument here has been that various memory systems are largely dissociable neurobiologically, they are, of course, not *entirely* so. Some commonalities should be at least briefly alluded to that might also have practical and ethical implications when considering the enhancement of human memory.

One such commonality is that many forms of memory involve the same three stages: encoding, storage, and retrieval. That is, information or skills must somehow be (i) absorbed or imprinted into memory in the first place; (ii) stored over time for later recall or expression; and finally, (iii) be retrieved or executed at some later time. Despite distal neuroanatomical locations, these stages may share similar neurobiological mechanisms across the various memory subtypes. In principle, then, a given drug (or other intervention) might be designed to have a broad or near-universal effect on one of these memory stages. The example of the 'encoding' stage is particularly illustrative: a huge body of research has shown that general level of arousal and attention strongly affects the success of encoding new information (Cahill & Alkire, 2003; Cahill, Gorski, & Le, 2003; LaBar & Phelps, 1998). This fact suggests that any intervention that could successfully modulate arousal could have general effects on all (or at least many) forms of memory – as may be the case with the pharmacological effects of caffeine, for instance (Nehlig, Daval, & Debry, 1992), or prescription stimulants (Smith & Farah, 2011; Wilens, 2006).

Second, we reiterate that the largely dissociable neural substrates of these various systems are a simplification of a complex and not-fully-understood neurobiological reality. Furthermore, even our present limited understanding suggests that certain forms of memory are inextricably intertwined at the neurobiological level: the encoding of both semantic and episodic memory, for instance, relies critically on medial temporal lobe structures, suggesting that it may be difficult to enhance or diminish one without simultaneously affecting the other. These caveats should be kept in mind when considering the otherwise striking dissociability of memory systems.

Relatedly, despite broad support, the multiple-system view of memory has not been without its critics (Stanley & Krakauer, 2013). The central critique of a systems view is the argument that evidence from neuropsychological lesion patients and other sources merely demonstrates semi-dissociable memory *processes*, but not necessarily distinct *systems* (Foster & Jelicic, 1999). Other critiques have focused on how specific forms of memory, such as *implicit* memory, are defined, casting many doubts on the ways in which the field operationalizes specific memory systems (Roediger III, 2003; Stanley & Krakauer, 2013); yet others have questioned the original rationale and justification behind the division of memory systems (Tulving, 2007). All in all, readers should be aware that although the multiple systems view of memory enjoys wide empirical and theoretical support, ultimately it remains an interpretation and model of neurobiological and psychological data rather than an established fact.

## 6. Conclusions and future directions

The prospect of memory enhancement raises ethical issues for a number of different stakeholders, including scientists who develop memory enhancements, physicians who may act as gatekeepers to their distribution, individuals who will choose to use (or not to use) memory enhancers, parents who are faced with the prospect of choosing for their children, employers and educators who will face new challenges in the management and evaluation of people who will enhance (or not enhance) their memories, regulatory agencies who may find their remit moving beyond therapy and into the enhancement world, and legislators and the public who will need to decide how to integrate the reality of memory enhancement into their worldviews.

We suggest that thinking about the ethical issues involved in enhancing 'memory' – conceived of as a single, monolithic concept or brain system – obfuscates many critical questions. For those engaging with these issues, both theoretical and empirical work should take into account the highly distinctive neurobiological and neurochemical systems that make up the panoply of human memory. Widespread public misperceptions and misunderstandings about the functioning and dysfunction of memory (Simons & Chabris, 2011) will need to be addressed as more and more individuals adopt an unregulated 'do-it-yourself' approach to memory enhancement (Fitz & Reiner, 2013); also critical will be empirical data on public perceptions of enhancement of different forms of memory, and what social and ethical concerns are key to each (Maslen et al., 2014). Does the public view enhancement of all kinds of memory as equally desirable, equally fair, or equally risky? For instance, one recent study suggested that people's views of narrative vs. working memory hardly differed at all on several dimensions, including how comfortable they were with enhancement of each memory type, or how much they thought enhancement of that kind of memory would change the person's identity (Cabrera et al., 2015). Unfortunately, very few studies to date bear on the question of public attitudes to various kinds of memory enhancement, and yet such information has large and obvious implications for any philosophical or ethical debate on these topics (Cabrera, Fitz, & Reiner, 2014; Cabrera et al., 2015). If people have no interest in strengthening autobiographical memory, for instance, there is little cause for extended debate on the implications of such an enhancement. Conversely, if there is wide interest in a given form of memory enhancement, then there is all the more reason for a rapid and thorough discussion of the risks and rewards of such an enhancement, and for developing sound policy to guide its use.

The multiplicity of memory enhancementLaxton, A. W., Tang‐Wai, D. F., McAndrews, M. P., Zumsteg, D., Wennberg, R., Keren, R., . . . Smith, G. S. (2010). A phase I trial of deep brain stimulation of memory circuits in Alzheimer's disease. *Annals of Neurology, 68*(4), 521-534.

Levy, N. (2007). *Neuroethics: Challenges for the 21st century*: Cambridge University Press.

Liao, S. M., & Sandberg, A. (2008). The normativity of memory modification. *Neuroethics, 1*(2), 85-99.

Lin, P., Mehlman, M., Abney, K., & Galliott, J. (2014). Super Soldiers (Part 1): What is Military Human Enhancement? *Global Issues and Ethical Considerations in Human Enhancement Technologies*, 119.

Loftus, E. F. (2013). 25 Years of Eyewitness Science…… Finally Pays Off. *Perspectives on Psychological Science, 8*(5), 556-557.

Loveman, E., Green, C., Kirby, J., Takeda, A., Picot, J., Payne, E., & Clegg, A. (2006). The clinical and cost-effectiveness of donepezil, rivastigmine, galantamine and memantine for Alzheimer's disease.

Lynch, G. (2002). Memory enhancement: the search for mechanism-based drugs. *Nature Neuroscience, 5*, 1035-1038.

Lynch, G., Palmer, L. C., & Gall, C. M. (2011). The likelihood of cognitive enhancement. *Pharmacology Biochemistry and Behavior, 99*(2), 116-129.

Maslen, H., Faulmüller, N., & Savulescu, J. (2014). Pharmacological cognitive enhancement—how neuroscientific research could advance ethical debate. *Frontiers in Systems Neuroscience, 8*.

McDonald, R. J., & White, N. M. (1993). A triple dissociation of memory systems: hippocampus, amygdala, and dorsal striatum. *Behavioral neuroscience, 107*(1), 3.

McGaugh, J. L. (2000). Memory--a century of consolidation. *Science, 287*(5451), 248-251.

Michaelian, K. (2011). Is memory a natural kind? *Memory Studies, 4*(2), 170-189.

Michaelian, K. (2015). Opening the doors of memory: is declarative memory a natural kind? *Wiley Interdisciplinary Reviews: Cognitive Science, 6*(6), 475-482.

Milner, B., Corkin, S., & Teuber, H.-L. (1968). Further analysis of the hippocampal amnesic syndrome: 14-year follow-up study of HM. *Neuropsychologia, 6*(3), 215-234.

Minzenberg, M. J., & Carter, C. S. (2008). Modafinil: a review of neurochemical actions and effects on cognition. *Neuropsychopharmacology, 33*(7), 1477-1502.

Moreno, J. D. (2003). Neuroethics: an agenda for neuroscience and society. *Nature Reviews Neuroscience, 4*(2), 149-153.

Moreno, J. D. (2006). *Mind wars: Brain research and national defense*: Dana Press Washington, DC.

Moscovitch, M., Rosenbaum, R. S., Gilboa, A., Addis, D. R., Westmacott, R., Grady, C., . . . Winocur, G. (2005). Functional neuroanatomy of remote episodic, semantic and spatial memory: a unified account based on multiple trace theory. *Journal of anatomy, 207*(1), 35-66.

Nehlig, A., Daval, J.-L., & Debry, G. (1992). Caffeine and the central nervous system: mechanisms of action, biochemical, metabolic and psychostimulant effects. *Brain Research Reviews, 17*(2), 139-170.

Norton, M. I., & Ariely, D. (2011). Building a better America—One wealth quintile at a time. *Perspectives on Psychological Science, 6*(1), 9-12.
*** **Article in press at** *Neuroethics* ***    32